\documentclass[12pt,fleqn]{article}
\begin{document}
\title{\bf A Model for Long Term Climate Change}
\author{\bf B G Sidharth and B S Lakshmi\footnote{Department of
Mathematics,JNTU College of Engineering
Kukatpally,Hyderabad.}\\\small Centre For Applicable Mathematics
and Computer Sciences\\\small B M Birla Science Center,Adarsh
Nagar,Hyderabad}
\date{}
\maketitle
\begin{abstract}
In this paper we consider the well-known cyclical model for
climate change ,for example from glaciation to the present day and
back, and compare this with a two-state model in Quantum
Mechanics.We establish the full correspondence.
\end{abstract}
\section{\bf Introduction}
In the Northern Hemisphere,the last glaciation took place 18,000
years ago.\cite{pri} The ice layer was several kilometers thick.It
covered up to the middle of US and Europe and right up to
Paris.Infact the current situation is one in which the continental
ice extends only up to Greenland and this was established 10,000
years ago.That is, the earth has in a few thousand years undergone
a tremendous transition between two completely different states in
a very short span of time-- geologically speaking. The earth can
be seen as a point object in space receiving solar radiation and
emitting infra-red radiation back into space.We consider the only
important state variable,the mean
 temperature T,with respect to time given by the heat balance equation
\begin{equation}
\frac{dT}{dt} = \frac{1}{C}(Q(1-a(T))-\varepsilon_{B} \sigma
T^{4})
\end{equation}

 $\sigma$ ,the Stefan constant $\varepsilon_{B}$,
emissivity factor, C,  the heat capacity of the system Q  the
Solar Constant and a, the albedo,which expresses the part of solar
radiation emitted back into space. The equation (1) admits two
steady states $T_{a}$,$T_{b}$, $T_{a}$, being  the  present day
climate and $T_{b}$, the glaciation time. A third state $T_{0}$ is
unstable and separates the above two stable states. As is well
known,in systems involving only one variable,U the kinetic
potential is given by \cite{ber}
\[ U = -\int dx F(x) \]
In the present case, the kinetic potential
 U(T) is given by \begin{equation}
 U(T) = -\frac{1}{C}\int dT (Q(1-a(T))-\epsilon_{B}\sigma T^{4})
 \end{equation}
Now climatic systems like any other physical system are
continuously subject to statistical
 fluctuations,the random deviation from deterministic behavior.
We include the effect of the fluctuations in a random force
F(t).The energy balance equation ,(1),and (2) now become a
stochastic differential equation of the form
\begin{equation}
\frac{dT}{dt} = -\frac{\partial U(T)}{\partial T} + F(T)
\end{equation}
The important new element introduced by this enlarged description
is that different states become connected through the
fluctuation.That is ,starting from some initial state the system
will reach any other state,sooner or later.This is true  for the
stable states $T_{a}$ and $T_{b}$ ,taken as initial states,which
become some sort of  meta-stable states.\\The time scale of this
phenomenon is determined
 by two factors:the potential barrier\[ \Delta U_{a,b} = U(T_{0})- U (T_{a,b}) \]
and the strength of the fluctuations  as measured by the variance ,$q^{2}$ of F(t)in (3).\\
The mean transition time from state $T_{a}$ or $T_{b}$ via the
unstable state $T_{0}$ is given by \cite{pri}
\begin{equation} T_{a,b} \sim
e^{\frac{\Delta U_{a,b}}{q^2}}
\end{equation}
\section{\bf The Model}
We will now model the above situation in terms of  a Quantum
Mechanical two state system.In the Quantum Mechanical world a two
state system\cite{fey} could be represented by the equations

\[i\hbar \frac{dC_{1}}{dt}= H_{11}C_{1} +H_{12}C_{2}\]
\begin{equation}
i\hbar \frac{dC_{2}}{dt}= H_{21}C_{1} +H_{22}C_{2}
\end{equation}
where the coefficients $H_{ij}$ are the Hamiltonian matrix and C is given by the vector, \\
\[C  \equiv  (C_{1},C_{2})\]We now identify C with T and its two
states $T_{a}$ and $T_{b}$ and write
 \[C  \equiv  (C_{1},C_{2}) \equiv (T_{a},T_{b})\equiv T \]
 This reduces (5) to
\begin{equation} i \hbar \frac{dT}{dt}= HT \end{equation} where H
is the 2x2 matrix. Taking $\bar{t}$ =$\frac{t}{i\hbar}$ reduces
(6) to
\begin{equation}
  \frac{dT}{d\bar{t}}= HT
\end{equation}
Reverting back to equation(3) and replacing t with T we have
\[ \frac{dT}{dt}=-\frac{\partial U(T)}{\partial t}+F(t) \]
Taking \[-\frac{\partial U(T)}{\partial t}+F(t)=H(T), \textup{we
get}
\]
\begin{equation}
  \frac{dT}{dt}= HT
\end{equation}
(8) is identical to (7).We note that from Quantum Mechanical
Theory,
\begin{equation} \hspace{10mm}\Delta t \hspace{10mm} \alpha \hspace{10mm} \frac{\hbar}{H}
\end{equation}
To proceed further and establish the correspondence fully,we now
observe $H_{12}$=$H_{21}$ is proportional to the transition
probability of $C_{1}$ to  $C_{2}$ i e., $T_{a}$ to $T_{b}$ In
fact\cite{rei} \textup {\[ H_{12}\hspace{5mm} \alpha
\hspace{5mm}e^{-U/kT}\]}  So the transition time in Quantum Theory
is given by
\begin{equation}
\Delta t \hspace{5mm}\alpha
\hspace{5mm}\frac{1}{H}\hspace{5mm}\alpha\hspace{5mm}e^{U/kT}
\end{equation}
 (10) can be identified with equation (4). This establishes the required
 result.\\
 The reduction of the Climate problem to the analogous Quantum
 Mechanical problem has interesting consequences which need to be
 studied further.This is all the more so because in recent years
 ``Scaled'' Quantum effects have been studied in macroscopic systems
 and even a Bode-Titius type relation for planetary distances has
 been deduced on the basis of ``quantized'' energy
 levels.\cite{bgs,bgg,car,agn}


\begin{thebibliography}{99}
\bibitem{pri}I.Prigogine,G.Nicolis,``Exploring\hspace{1mm}Complexity",R.Piper,GmbH,Munich,1989,pp.226 ff.
\bibitem{ber}A.Berger.,(ed)``Climatic Variations and Variability:facts and theories'',Reidel,Dordrecth,1981.
\bibitem{fey}R.P.Feynman,R.B.Leighton,and\hspace{1mm}M.Sands,``The\hspace{1mm}Feynman\hspace{1mm}Lectures\hspace{1mm}on\hspace{1mm}Physics'',Vol.III,Addison-Wesley\hspace{1mm}PublishingCo.,Inc.,Massachussets,1965,Chapter 8 ff.
\bibitem{rei}F.Reif,``Statistical and Thermal Physics'',McGraw-Hill,Singapore,1965.
\bibitem{bgs}B.G.Sidharth,Chaos,Solitons and Fractals,12,2001,613-616.
\bibitem{bgg}B.G.Sidharth,Chaos,Solitons and Fractals,12,2001,1371-1373.
\bibitem{car}L.Nottale,``Fractal Space-Time and Microphysics:Towards a Theory of Scale Relativity'',World Scientific,Singapore,1993,p.312.
\bibitem{agn}A.G.Agnese and R.Festa,Phys.Lett.A.,227,1997,p.165-171.
\end{thebibliography}
\end{document}